\documentstyle[psfig,multicol,aps]{revtex}
\begin{document}
\draft
\title{Pion Scalar Form Factor and the Sigma Meson}
\author{Torben Hannah\cite{THannah}}
\address{Nordita, Blegdamsvej 17, DK-2100 Copenhagen {\O}, Denmark}
\maketitle
\begin{abstract}
From recent analysis of the $\pi\pi$ scattering amplitude, it
has been claimed that there exists a broad and light $\sigma$ meson.
However, if this meson really exists, it must also
appear in other observables such as the pion scalar form factor.
With the use of unitarity and dispersion relations together with
chiral perturbation theory, this form factor is analyzed in the
complex energy plane. The result agrees well with the empirical
information in the elastic region and reveals a resonance pole
at $\sqrt{s}=445-i235$ MeV. This gives further strong evidence for
the existence of the $\sigma$ meson.
\end{abstract}

\begin{multicols}{2}

It is still controversial whether a broad and light isosinglet scalar
meson really exists. After having been omitted from the
Particle Data Group for more than twenty years, this scalar meson
has reappeared in the last two editions of the Particle Data Group
under the entry $f_0$(400-1200) or $\sigma$ \cite{ref:PDG98}. This
reintroduction is based upon several new theoretical analyses
\cite{ref:TR96,ref:Ishida97,ref:KLL97}, which have given support
to the existence of a broad and light scalar meson. In addition,
there has also recently been some experimental indications of
$\sigma$ production from central $pp$ collisions \cite{ref:Alde97}.  

The theoretical evidence for the existence of the $\sigma$ meson comes
mainly from model-dependent analysis of $\pi\pi$ scattering
\cite{ref:TR96,ref:Ishida97,ref:KLL97}. However, if this
meson in fact exist, it must also appear in other processes containing
$\pi\pi$ in the final state. Therefore, in order to further investigate
whether the $\sigma$ meson really exists, other processes should also be
analyzed. The pion scalar form factor is such a process which can
give further important information on the $\sigma$ meson. This is somewhat
similar to the well-known case of the $\rho$(770) meson, where the
information on this resonance can be obtained both from $\pi\pi$ scattering
and from the pion vector form factor.

The pion scalar form factor has previously been calculated using the
inverse amplitude method (IAM) \cite{ref:Han97,ref:Tru88}. This
method is based upon the combination of unitarity and dispersion relations
together with chiral perturbation theory (ChPT) \cite{ref:We79,ref:GL84}
and has been used in order to extend the range of applicability of ChPT.
In particular, the IAM has been applied in order to account for possible
resonances since this method, contrary to ChPT, can produce resonance poles
in the complex energy plane. In this brief report, the scalar form factor
is investigated further using the IAM and the result is analyzed in the
complex energy plane in order to find possible resonance poles corresponding
to the $\sigma$ meson. 

The pion scalar form factor $F$ is given by the matrix element of the
quark density
\begin{equation}
\label{eq:defF}
\langle\pi^i(p_2)|\bar{u}u+\bar{d}d|\pi^j(p_1)\rangle = \delta^{ij}F(s) ,
\end{equation}
where $s=(p_2-p_1)^2$. This form factor is analytical in the complex $s$
plane with a unitarity cut starting at the $\pi\pi$ threshold. In the
elastic region the unitarity relation is given by
\begin{equation}
\label{eq:uni}
{\rm Im}F(s) = \sigma (s)F^{\ast}(s)t^0_0(s) ,
\end{equation}
where $\sigma (s)$ is the phase-space factor and $t^0_0$ is the isosinglet
scalar $\pi\pi$ partial wave. This relation implies that the phase of
$F$ will coincide with the $\pi\pi$ phase shift $\delta^0_0$ in accordance
with Watson's final-state theorem \cite{ref:Wat54}. The first important
inelastic effect starts at around 1 GeV and is due to the
$K\bar{K}$ intermediate state. Since the main interest will be in
energies well below this inelastic effect, in the following the form factor
will only be calculated using the elastic approximation.

The scalar form factor has been calculated to two loops in ChPT both
by a dispersive analysis \cite{ref:GM91} and more recently by a full
field theory calculation \cite{ref:BCT98}. The result can be written as
\begin{equation}
\label{eq:ChPTF}
F(s) = F^{(0)}(s)+F^{(1)}(s)+F^{(2)}(s) ,
\end{equation}
where $F^{(0)}$ is the leading order result, $F^{(1)}$ the
one-loop correction, and $F^{(2)}$ the additional two-loop
correction. Since the form factor will be normalized to $F(0)=1$
in the following, one has for the leading order term $F^{(0)}=1$.
The one-loop correction is given in terms of the single one-loop low-energy
constant $l^r_4$ \cite{ref:GL84}, whereas the two-loop correction
contains the additional one-loop low-energy constants $l^r_1$, $l^r_2$,
and $l^r_3$ together with the two-loop low-energy constants
$r^r_{S2}$ and $r^r_{S3}$ \cite{ref:BCT98}. The superscript $r$
indicates that these low-energy constants depend on the renormalization
scale $\mu$, whereas the full form factor is scale-independent.
Since ChPT is a perturbative expansion, the unitarity relation
(\ref{eq:uni}) will only be satisfied perturbatively
\begin{eqnarray}
\label{eq:puni}
{\rm Im}F^{(0)}(s) & = & 0 , \nonumber \\
{\rm Im}F^{(1)}(s) & = & \sigma (s)t^{0(0)}_0(s) , \nonumber \\
{\rm Im}F^{(2)}(s) & = & \sigma (s) \left[ {\rm Re}F^{(1)}(s)
t^{0(0)}_0(s)+{\rm Re}t^{0(1)}_0(s) \right] .
\end{eqnarray}
This perturbative unitarity will restrict the applicability of ChPT to the
very low-energy region. However, with the use of the IAM,
the range of applicability of ChPT can be substantially extended
to also include resonance regions. The starting point for
this method is to write down a dispersion relation for the inverse
of the form factor $\Gamma =1/F$ \cite{ref:Han97,ref:Tru88}. In this
dispersion relation the unitarity relation (\ref{eq:uni}) gives
${\rm Im}\Gamma =-{\rm Im}F/|F|^2=-\sigma t/F$. Expanding this quantity
to two loops in ChPT gives  
${\rm Im}\Gamma =-\sigma [t^{(0)}(1-{\rm Re}F^{(1)})+{\rm Re}t^{(1)}]$
which can be used on the unitarity cut in the dispersion relation for $\Gamma$.
The subtraction constants may also be evaluated by expanding the function
$\Gamma$ to two-loop order as $\Gamma^{(2)}=1-F^{(1)}+{F^{(1)}}^2-F^{(2)}$.
Thus, neglecting possible zeros in the form factor, one has the following
dispersion relation
\begin{eqnarray}
\label{eq:disp1}
\frac{1}{F(s)} & = & 1+a_1s+a_2s^2-\frac{s^3}{\pi}
\int^{\infty}_{4M^2_{\pi}}ds' \sigma (s')\times \nonumber \\
&& \frac{t^{0(0)}_0(s')\left[ 1-{\rm Re}F^{(1)}(s')\right]
+{\rm Re}t^{0(1)}_0(s')}{s'^3(s'-s-i\epsilon )} ,
\end{eqnarray}
where three subtractions are used in order to make the dispersion integral
convergent. This relation can be simplified by writing a dispersion relation
for the function $\Gamma^{(2)}$. Using perturbative unitarity (\ref{eq:puni})
one finds the this dispersion relation will be exactly similar to the one
given in Eq. (\ref{eq:disp1}).
Thus, the IAM to two loops in the chiral expansion gives the
form factor as \cite{ref:Han97}
\begin{equation}
\label{eq:IAM2}
F(s) = \frac{1}{1-F^{(1)}(s)+{F^{(1)}}^2(s)-F^{(2)}(s)} .
\end{equation}
This expression for the form factor is formally equivalent to
the [0,2] Pad\'{e} approximant applied on ChPT and will therefore
coincide with the chiral expansion up to two loops. However, with
the IAM the range of applicability of ChPT is substantially extended.
This is based upon the fact that the expansion of $t/F$ used in
the IAM works well over a much larger region than the
corresponding expansion of $F^{\ast}t$ used in ChPT. In fact,
the former expansion works well throughout the elastic region,
even when the form factor has a resonant character \cite{ref:Tru88}.

However, the IAM may generate poles on the physical sheet
which violate the analyticity requirement. These
poles are caused by the high-energy part of the dispersion
integral in Eq. (\ref{eq:disp1}). Since this part is not
expected to be well approximated, it may cause the right-hand
side of Eq. (\ref{eq:disp1}) to vanish and thereby generate spurious
poles in the form factor. These poles should in principle be
removed without any significant influence in the region of applicability
of the IAM. A rather general method to remove possible poles and thereby
restore analyticity is to put the imaginary part of the IAM
back into a dispersion relation.
With three subtractions the result can be written as
\begin{equation}
\label{eq:polIAM2}
F(s) = 1+\mbox{$\frac{1}{6}$}\langle r^2\rangle s+cs^2
+\frac{s^3}{\pi}\int^{\infty}_{4M^2_{\pi}}\frac{{\rm Im}F(s')ds'}
{s'^3(s'-s-i\epsilon )} ,
\end{equation}
where both the subtraction constants and ${\rm Im}F$ are calculated
from the IAM (\ref{eq:IAM2}). Here, it is assumed
that ${\rm Im}F$ does not contain any poles on the unitarity cut
and three subtractions
are used in order to suppress the high-energy part of the
dispersion integral. Without any poles Eq. (\ref{eq:polIAM2}) is just
an identity, but with poles the output will in general be different
from the input. However, in the region where the IAM is applicable,
the difference between the form (\ref{eq:polIAM2}) and the form
(\ref{eq:IAM2}) should be small. In fact, this method to remove
possible poles is equivalent to the subtraction of the poles on the
physical sheet from the original IAM (\ref{eq:IAM2}). This will be
discussed in more detail elsewhere \cite{ref:HT98}.

\begin{figure}
\centerline{\psfig{figure=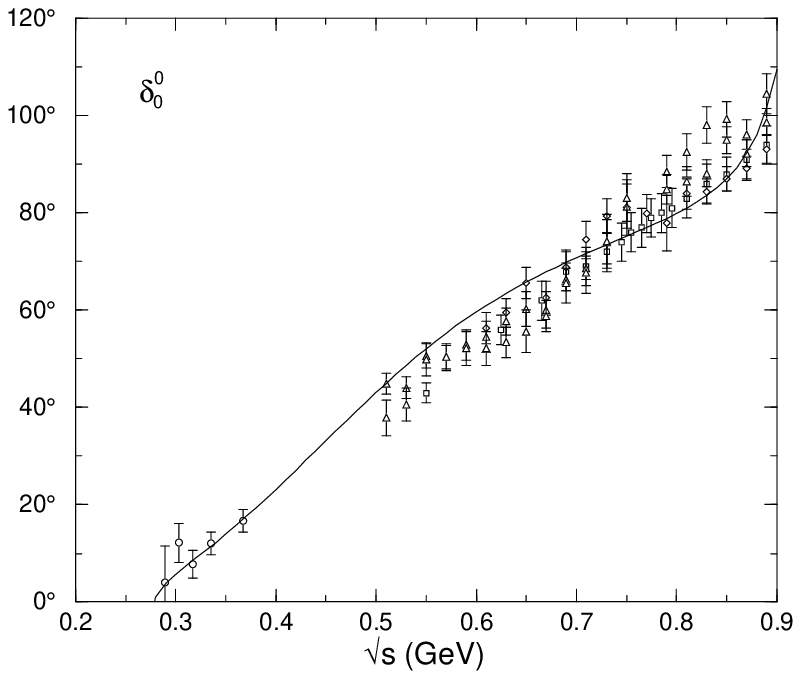,width=9cm,angle=-90}}
\caption{The phase $\delta^0_0$ of the scalar form factor in the
elastic region below $\protect\sqrt{s}=0.9$ GeV. The experimental $\pi\pi$
phase shifts are from Ref. \protect\cite{ref:Ros77} (circles), Ref.
\protect\cite{ref:Pro73} (squares), Ref. \protect\cite{ref:Hyams73}
(diamonds), and Ref. \protect\cite{ref:EM74} (triangles).}
\label{Fig1}
\end{figure}

The IAM to two loops depends on a number of
low-energy constants which have to be determined phenomenologically.
Unfortunately, the scalar form factor is not directly accessible
to experiment. However, in the elastic region the phase of $F$ is
given by the $\pi\pi$ $I=0$ $S$ phase shift $\delta^0_0$, which is
known experimentally. Fitting these phase shifts up to 0.9 GeV
\cite{ref:Ros77,ref:Pro73,ref:Hyams73,ref:EM74} and using the value of the
pion scalar radius $\langle r^2\rangle =0.60$ ${\rm fm}^2$
\cite{ref:GM91,ref:DGL90}, some of the
low-energy constants in the IAM to two loops have previously been determined
without taking possible poles into account \cite{ref:Han97}. Here,
in order to remove spurious poles, this fit is repeated with the form
factor given by Eq. (\ref{eq:polIAM2}). The result is shown in
Fig. \ref{Fig1} together with the experimental $\pi\pi$ phase shifts,
from where it is observed that the IAM agrees rather well with the main bulk
of the data all the way up to 0.9 GeV. Thus, the IAM satisfies
Watson's final-state theorem \cite{ref:Wat54} quite well in the whole
elastic region. This fit gives the following values for the
low-energy constants
\begin{eqnarray}
\label{eq:lec}
l^r_4 & = & 1.53\times 10^{-3} ,\nonumber \\
r^r_{S2} & = & 2.25\times 10^{-3} ,\nonumber \\
r^r_{S3} & = & 7.60\times 10^{-5}
\end{eqnarray}
at the renormalization scale $\mu =M_{\rho}=770$ MeV. Since the experimental
data are not very consistent with each other, there has not been
assigned any error bars on these low-energy constants. However, the obtained
values of $l^r_4$ and $r^r_{S3}$ agree rather well with the recent
determination of these low-energy constants using two-loop ChPT
\cite{ref:BCT98}. As for $r^r_{S2}$, this low-energy constant has
so far only been estimated on the basis of the resonance saturation
hypothesis \cite{ref:Eck89} with a result \cite{ref:BCT98} that is somewhat
smaller than the value obtained above. In the future, it might be possible
to determine this two-loop low-energy constant from independent observables
\cite{ref:Han96} and thereby check the value obtained here.

The scalar form factor can be defined in the whole complex
$s$ plane. Since it contains cuts starting at the $\pi\pi$
threshold, this will involve different Riemann
sheets. In the elastic approximation there are two Riemann sheets,
which are defined according to the sign of the center of mass momenta
$q=\sqrt{s-4M^2_{\pi}}/2$. The first or physical sheet has positive
values of ${\rm Im}q$, whereas the second or unphysical sheet has
negative values of ${\rm Im}q$. The form factor given by either
Eq. (\ref{eq:IAM2}) or Eq. (\ref{eq:polIAM2}) can indeed be extended
analytically to the whole complex $s$ plane. This analytic continuation
will involve infinitely many Riemann sheets since the cut in the IAM
comes from logarithmic functions. However, only two of these sheets
correspond to the first and second Riemann sheet that the form factor
should reproduce.

\begin{figure}
\centerline{\psfig{figure=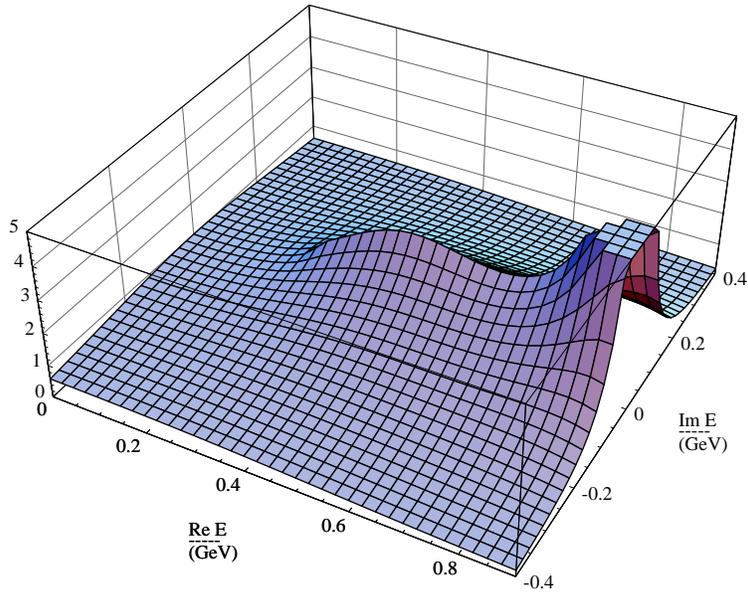,width=10cm}}
\caption{The scalar form factor $|F|^2$ in the complex energy plane
on the first Riemann sheet. The complex energy $E$ is defined according
to $s=({\rm Re}E+i{\rm Im}E)^2$.}
\label{Fig2}
\end{figure}

In Fig. \ref{Fig2} the absolute square of the form factor
(\ref{eq:polIAM2}) is shown in the complex energy plane on the first
Riemann sheet. On the real axis the result agrees very well with the
result of a dispersive analysis \cite{ref:GM91,ref:DGL90}, where the
scalar form factor has been determined from the experimental
$\pi\pi /K\bar{K}$ phase shifts. Furthermore, the form factor
(\ref{eq:polIAM2}) is analytic in the whole complex energy plane
with the correct cut structure starting at the $\pi\pi$ threshold.
This is different from the original IAM (\ref{eq:IAM2}) which
generates a pole on the negative $s$ axis. However, this pole is
removed by using the form (\ref{eq:polIAM2}) without any significant
influence on the result in the region shown in Fig. \ref{Fig2}.

\begin{figure}
\centerline{\psfig{figure=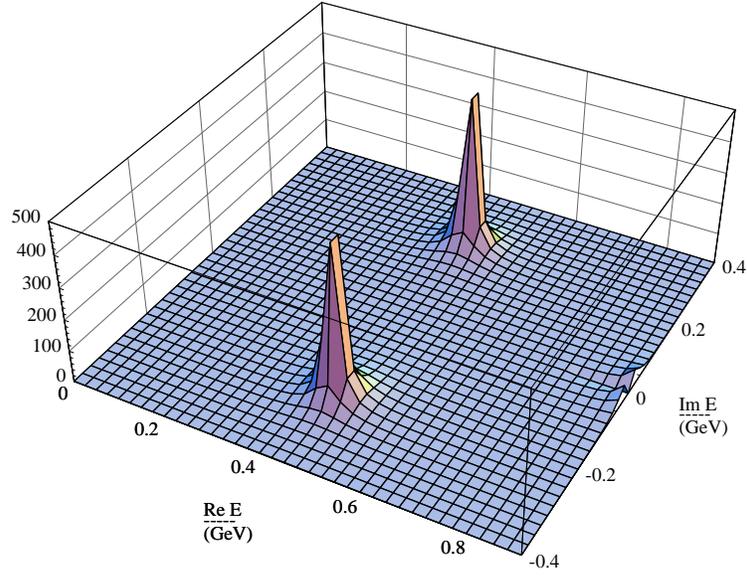,width=10cm}}
\caption{The scalar form factor $|F|^2$ in the complex energy plane
on the second Riemann sheet.}
\label{Fig3}
\end{figure}

From this figure it is also observed that around $0.4-0.5$ GeV the form
of $|F|^2$ is somewhat reminiscent of a resonant structure. However,
in order to investigate whether this form is really associated with
a resonance, one has to consider the second Riemann sheet. On this sheet
resonances are characterized by poles in the complex energy plane,
where the mass ($M_R$) and width ($\Gamma_R$) of the resonance can be
related to the position of the pole by
\begin{equation}
\label{eq:resonance}
\sqrt{s_{pole}} = M_R-i\frac{\Gamma_R}{2} .
\end{equation}
In Fig. \ref{Fig3} the absolute square of the form factor (\ref{eq:polIAM2})
is shown in the complex energy plane on the second Riemann sheet. One finds
that $|F|^2$ indeed generates two complex conjugated poles corresponding to
a broad and light resonance. In fact, the position of these poles is the same
for the two expressions of the form factor given by Eq. (\ref{eq:IAM2}) and
Eq. (\ref{eq:polIAM2}), respectively. From the position of the pole, the
mass and width of this $\sigma$ meson is given by
\begin{equation}
\label{eq:mwsigma}
M_{\sigma} = 445\;{\rm MeV}\;\; ,\;\;\Gamma_{\sigma} = 470\;{\rm MeV} .
\end{equation}
This compares rather well with the values $M_{\sigma}=470$ MeV
and $\Gamma_{\sigma}=500$ MeV obtained in Ref. \cite{ref:TR96}.
However, rather different values for $M_{\sigma}$ and $\Gamma_{\sigma}$
have also been obtained with other theoretical models
\cite{ref:Ishida97,ref:KLL97}. Therefore, it is important to reduce
the model-dependence when the mass and width of the $\sigma$ meson are
determined. The IAM is based solely on the use of unitarity and
dispersion relations together with ChPT. Therefore, within
this approach, any model-dependence in the mass and width of
the $\sigma$ meson is due to higher order terms in
the chiral expansion together with the present uncertainties in the
values of the low-energy constants.

The IAM is in fact a systematic approach which can be applied to any
given order in the chiral expansion. Originally, this method was applied
to the scalar form factor in the one-loop approximation \cite{ref:Tru88}
with a result that is formally equivalent to the [0,1] Pad\'{e} approximant
applied on ChPT. In this case the IAM depends on the single one-loop
low-energy constant $l^r_4$ which can be determined from the ratio
$F_K/F_{\pi}$ \cite{ref:GL84,ref:Han96}. With this low-energy constants
fixed, the IAM to one loop agrees rather well with the empirical information
up to about 0.5 GeV. The one-loop approximation also contains a pole on the
negative $s$ axis. However, this pole can be removed by using the method
discussed previously without any significant influence on the result in the
elastic region. Extending this result to the whole complex energy plane,
one finds that the IAM to one loop also generates a resonance pole,
where the corresponding mass and width of this resonance are given by
\begin{equation}
\label{eq:mwsigma1}
M_{\sigma} = 463\;{\rm MeV}\;\; ,\;\;\Gamma_{\sigma} = 393\;{\rm MeV} .
\end{equation}
Comparing these values with the values obtained from the IAM to two loops
(\ref{eq:mwsigma}), it is observed that the masses are very similar, whereas
the difference in the widths is somewhat larger. However, this difference
is not significant compared to the large uncertainty in the width
of the $\sigma$ meson given
by the Particle Data Group \cite{ref:PDG98}. In view of this the convergence
of the IAM is satisfactory for both the mass and width of the $\sigma$ meson.
Hence, the corrections to the values given in (\ref{eq:mwsigma})
due to even higher orders in the chiral expansion are expected to be of little
importance. There is also an uncertainty in the obtained values for $M_{\sigma}$
and $\Gamma_{\sigma}$ coming from the uncertainties in the values of the
low-energy constants. However, this effect should also be rather small
and could be estimated when the values of the low-energy constants in
the IAM are determined more accurately. 

The IAM is not restricted to the scalar form factor but this method is
in fact quite general and has been applied to other processes as well.
In particular, the IAM has been applied to $\pi\pi$ scattering where it
also generates a resonance pole corresponding to the $\sigma$ meson
\cite{ref:DP97}. In fact, the mass and width of the $\sigma$ resonance
obtained from $\pi\pi$ scattering, $M_{\sigma}=440$ MeV and
$\Gamma_{\sigma}=490$ MeV, agree very well with the values
obtained in (\ref{eq:mwsigma}). This strongly supports the consistency
of the IAM and gives additional evidence for the existence of the
$\sigma$ meson.

To summarize, the pion scalar form factor has been calculated
by the use of unitarity and dispersion relations together
with the chiral expansion. In order to satisfy the analyticity
requirement, possible poles on the physical sheet
are removed from this IAM. The result
agrees well with both the experimental $\pi\pi$ phase shifts and
a dispersive analysis in the whole elastic region. Making an analytic
continuation of the scalar form factor to the complex energy plane,
one finds a resonance pole corresponding to a broad and light
scalar meson. The values for the mass and width of this $\sigma$
meson are obtained in a rather model-independent way, contrary to
previous determinations where different theoretical models were
applied. Indeed, any model-dependence in the obtained values
for $M_{\sigma}$ and $\Gamma_{\sigma}$ should be rather small
compared to the present uncertainty in these quantities.
All this gives further strong evidence for the existence
of the controversial broad and light $\sigma$ meson.

\end{multicols}

\newpage

\vspace*{2cm}

\newpage

\vspace*{2cm}

\end{document}